\documentclass[preprint,prd, nofootinbib]{revtex4-1}
\usepackage{color}
\usepackage{graphicx}
\usepackage{epsfig}
\usepackage{subfig}
\usepackage{lipsum}
\usepackage{ctable}
\usepackage{hyperref}
\usepackage{cleveref}
\crefformat{footnote}{#2\footnotemark[#1]#3}
\usepackage{physics}
\usepackage{graphicx}% Include figure files
\usepackage{dcolumn}% Align table columns on decimal point
\usepackage{bm}% bold math

\begin{document}

	\title{ Compact  Stars in    the Non-minimally Coupled Electromagnetic Fields to Gravity}
	% Force line breaks with \\

	\author{ \"{O}zcan SERT}
	\email{osert@pau.edu.tr}
	\affiliation{Department of Mathematics, Faculty of Arts and Sciences,
		Pamukkale
		University,  20070   Denizli, T\"{u}rkiye}
	%\author{Muzaffer ADAK}
	%\email{madak@pau.edu.tr}
	%\affiliation{Department of Physics,
	%Pamukkale
	%University, 20070,  K{\i}n{\i}kl{\i},  Denizli, Turkey}


	\date{\today}% It is always \today, today,

	\begin{abstract}

		\noindent
		
%		In recent cosmological observations direction, the  modification of Einstein gravity   leads to the modification of Einstein-Maxwell gravity which  is the minimal theory with the minimal coupling  between electromagnetic fields and gravitation in $R+ F^2$ form. 

% In the context of   a compact star which have   high energy density, pressure, gravitational and electromagnetic fields,  the  non-minimal couplings   such as $Y(R)F^2$ type can be appeared, where. 

 We investigate the  gravitational  models with   the non-minimal  $Y(R)F^2$ coupled electromagnetic fields to gravity, in order to describe charged compact stars, where
  $Y(R) $ denotes a function of the Ricci curvature scalar  $R$   and $F^2$ denotes  the Maxwell  invariant term.
We
determine two parameter family of exact spherically symmetric static solutions  and the corresponding non-minimal model  
  without  assuming  any relation between   energy density of matter and pressure.   
   We give the mass-radius, electric charge-radius ratios and surface gravitational redshift which are obtained by the boundary  conditions. We reach  a wide range of possibilities  for the parameters $k$ and $\alpha$ in these solutions.   Lastly  we show that the models  can describe the compact stars even in the  more simple  case  $\alpha=3$.

		%\begin{description}

		% \item[PACS numbers]

		%\end{description}
	\end{abstract}
	
	\pacs{Valid PACS appear here}% PACS, the Physics and Astronomy
	\maketitle
	
	%\tableofcontents

	\def\ba{\begin{eqnarray}}
	\def\ea{\end{eqnarray}}
	\def\w{\wedge}

	%\documentclass[landscape]{slides}
	%\usepackage{xcolor}
	%\usepackage{amsfonts}
	%\usepackage{amsmath}
	%\usepackage{amsbsy}
	%\usepackage{amssymb,latexsym}
	%\special{landscape}
	
	%\begin{document}
	%\def\ba{\begin{eqnarray}}
	%\def\ea{\end{eqnarray}}
	%\def\w{\wedge}

\section{Introduction}

\noindent

Spherically symmetric solutions in gravity 
are fundamental tools  in order to describe 
the structure and physical properties  of compact stars.
There is a large number of interior exact spherically symmetric solutions of Einstein's theory  of gravitation (for reviews see \cite{kramer,delgaty}).
But, very few of them satisfy the necessary  physical and continuity  conditions for  a compact fluid.
% A huge number of analytical solutions of the Einstein gravitational field equations describing the interior structure of the static fluid spheres were found in the past 100 years (for reviews of the interior solutions of the Einstein gravitational field equations see [38?40])
Some of them  were given by Mak and Harko \cite{mak-harko-2005,mak-2013,harko2016} for an isotropic neutral spherically symmetric matter distribution.

A charged  compact star  may  be more stable \cite{Stettner}  and prevent the gravitationally collapse   \cite{Krasinski,Sharma}, 
therefore  it is interesting to   consider the case with charge. 
The charged solutions of Einstein-Maxwell field equations    which describe a strange quark star  were  found by Mak and Harko  \cite{mak-harko-2004} considering a symmetry of conformal motions with   MIT bag model.

Since the Einstein's theory of gravity has  significant observational problems at large  cosmological scales \cite{Overduin,Baer,Riess,Perlmutter,Knop,Amanullah,Weinberg,Schwarz},  one need  to search new  theories of gravitation which are  acceptable even at these scales.   
Some compact star solutions in such  modified theories as the hybrid metric-Palatini gravity \cite{danila-harko}  and  Eddington-inspired
Born-Infeld (EIBI) gravity \cite{harko-lobo-2013}, were studied numerically.

One can consider that a charged astrophysical object can described by  the minimal coupling between the gravitational and electromagnetic fields known as Einstein-Maxwell theory. However, 
the above problems of Einstein's gravity at large scales  can also lead to investigate  the $Y(R)F^2$ type modification   of Einstein-Maxwell theory
\cite{AADS,dereli3,Sert12Plus,Sert13MPLA,bamba1,bamba2,Sert-Adak12,dereli4,Turner,Mazzi,Campa}.
Such non-minimal  modifications  can  also be found  in \cite{AADS,dereli3,Sert12Plus,Campa,Prasanna,Drummond,Kunze,dereli4,bamba1,Horndeski,Mueller,Buchdahl2,Buchdahl,Sert13MPLA,bamba1,bamba2,Sert-Adak12,Dereli901,Turner,Mazzi,Sert2017,Baykal,Sert16regular,dereli2}  to obtain more information on  the interaction between electromagnetic and gravitational fields  and all  other energy forms. The non-minimal couplings
also can  arise  in   such compact objects
as black holes, quark stars and neutron stars which have very  high density gravitational and electromagnetic fields \cite{Sert2017}. If the extreme situations are disappeared, that is far from the compact stars   the model  turns out to be the Einstein-Maxwell case.

Here we consider the non-minimal  $Y(R)F^2$ type modification to the Einstein-Maxwell theory
and  generalize the exact solutions  for  radiation fluid case $k=1$  in \cite{Sert2017}   to the cases   with $k\neq 1$, inspired by the study  \cite{mak-harko-2004}.  We obtain the inner region solutions and construct the corresponding model which turns to the Einstein-Maxwell theory  in the outer region. We note   that  the inner  solutions  
recover   the solution obtained by Misner and Zapolsky  \cite{misner} for   charge-less case
and $b=0$.
We find the surface  gravitational redshift,  matter mass, total mass and charge in terms of boundary radius and the parameters $k$ and $\alpha$ via the continuity and boundary conditions.

We organize the paper  as follows:  
In section II, we give the non-minimal  $Y(R)F^2$ gravity model   and   field  equations   in order to  describe a compact fluid. 
 In section III, we obtain exact static, spherically  symmetric solutions of the model under the conformal symmetry and the corresponding $Y(R)$ function.
 In section IV,  we determine the total mass, charge and gravitational redshift of the star in terms of boundary radius and the parameters of the model $k$ and $\alpha$. We summarize the results in the last section.

\section{ The  Model  for a Compact Star} \label{model}

 Compact stars have very intense energy density,  pressure and gravitational fields.  They also  can have  very high electric fields 
 in order to balance the huge gravitational pulling  \cite{Stettner,Sharma,Krasinski,Felice1999,Felice1995,Anninos2001,Zhang1982,Yu2000,Bekenstein1971}.
 Even if they collapse, very high electric fields necessary to explain the formation of   electrically charged black holes \cite{Ray2003,Malheiro}. Moreover they can  have very strong magnetic fields \cite{Turolla}.
Therefore the new non-minimal interactions   between these electromagnetic and gravitational fields  with $Y(R)F^2$ type can arise  under the extreme situations. 
Thus we consider the following  non-minimal model for compact stars, which involves  the Lagrangian of  the electromagnetic source  $A\wedge J$ and the   matter part $L_{mat}$ in the action  \cite{Sert2017}
\begin{equation}\label{action}
I(e^a,{\omega_a}^b, A)  = \int_M{\{   \frac{1}{2\kappa^2} R*1 -Y(R) F\wedge *F + 2A\wedge J + L_{mat} + \lambda_a\wedge T^a  \} }\;.
\end{equation}
Here   $e^a$ is the orthonormal co-frame 1-form,  $ {\omega^a}_b$ is the Levi-Civita connection 1-form obtaining by the relation $T^a =  de^a + \omega^{a}_{\;\;b} \wedge e^b=0$, F is the electromagnetic tensor 2-form which is derived from   the electromagnetic potential $A$ via the exterior derivative, that is $ F=dA$;  $\lambda_a$ is the Lagrange multiplier which gives torsion-less space-times, $T^a=0$; $R$ is   the Ricci scalar  and  $J$  is  the electromagnetic current density        in   the star. 

   The co-frame variation of the action (\ref{action}) is given  by the following gravitational field equation   after eliminating the connection variation,
\begin{eqnarray}
\label{gfe}
- \frac{1}{2 \kappa^2}  R_{bc}
\wedge *e^{abc}  & = &   Y  (\iota^a F \wedge *F - F \wedge \iota^a *F)   +  Y_R F_{mn} F^{mn}*R^a 
\nonumber
\\
&& +   D [ \iota_b\;d(Y_R F_{mn} F^{mn} )]\wedge *e^{ab} + \  (\rho +p )u^a*u + p*e^a
\  ,
\end{eqnarray}
where we have used  the velocity 1-form    $u=u_a e^a $  for    time-like inertial observer  and  $ \frac{dY}{dR} = Y_R $.  Furthermore we have considered the matter  energy momentum tensor  has the energy density $\rho$ and the pressure $p$ as diagonal elements for the isotropic matter in the star.     
It is worth to note that  the total energy-momentum tensor which is right hand side of  the gravitational field equation  (\ref{gfe})    satisfies the conservation  relation \cite{Sert2017}. 
The electromagnetic potential variation of the action gives the following modified Maxwell field equation
\begin{eqnarray}
d(*Y F) &=& J \; .  \ \label{maxwell1}
\end{eqnarray} 
We have also the identity $ dF = d(dA)= 0 $. We will find  solutions to the field equations  (\ref{gfe}) and (\ref{maxwell1})  under the condition  \begin{eqnarray}\label{cond0}
Y_R F_{mn} F^{mn} = - \frac{ k}{ \kappa^2}  \label{YRFk} 
\end{eqnarray}
which eliminates the instabilities of the higher order derivatives in the theory. 
Here we note that the constant $k$ \footnote{Here we replaced $K$ in \cite{Sert2017} with $-k $ to continue with $k>0$ .}   determines the strength of the non-minimal coupling 
 between gravitational and  electromagnetic fields. 
 The case with    $k=0$  leads to $Y(R)= constant $ and this case corresponds to  minimal Einstein-Maxwell theory which can be considered as the exterior vacuum   Reissner-Nordstrom solution with $R=0$.
% If we consider  $Y(R) = R$  which give $F_{mn}F^{mn} =constant$ then this model has a constant electromagnetic field everywhere in the spacetime. 
The additional  features of the constraint (\ref{cond0}) can be found in \cite{Sert2017}.
On the other hand, the trace of the non-minimally coupled  gravitational field equation (\ref{gfe})  gives 
\begin{eqnarray}
\frac{1-k}{\kappa^2} R*1 = (\rho  - 3p)*1 \;. \label{trace}
\end{eqnarray}

Since the case with   $k= 1$ or $\rho= 3p $ is investigated as the  radiation fluid stars   in \cite{Sert2017},  we concentrate on the case with $k\neq 1$ or $\rho \neq 3p$ in this study.

\section{SPHERICALLY SYMMETRIC SOLUTIONS UNDER CONFORMAL SYMMETRY}

\noindent We take the following static, spherically symmetric   metric  and   Maxwell  2-form with the  electric component $E$  which has only radial dependence 
\begin{eqnarray}\label{metric}
ds^2 & =& -f^2(r)dt^2  +  g^2(r)dr^2 + r^2d\theta^2 +r^2\sin^2\theta d \phi^2 \;,
\\
\label{electromagnetic1}
F     &= & E(r) e^1\wedge e^0 .
\end{eqnarray}
Then  the  charge in  the star   can be obtained from the integral of the current density 3-form $J$ over the volume $V$ with radius $r$
using  the Maxwell  equation (\ref{maxwell1}) 
\begin{eqnarray}\label{q1}
q(r) = \frac{1}{4\pi}\int_V J =\frac{1}{4\pi} \int_{V} d*YF  = YEr^2\;.
\end{eqnarray}
The Ricci scalar for the metric (\ref{metric}) is calculated as
\begin{eqnarray}\label{Ricci}
R = \frac{2}{g^2} \left(   \frac{f' g' }{fg}  -   \frac{f''}{f}  +   \frac{2g'}{gr} -  \frac{2f'}{fr}  +  \frac{g^2 -1 }{r^2}   \right) \;.
\end{eqnarray}

For $k\neq 1 $ or $\rho \neq 3p $ the gravitational field equation (\ref{gfe}) gives 
\begin{eqnarray}
\frac{1}{\kappa^2 g^2}(    \frac{2g'}{rg}  + \frac{g^2-1}{r^2}  )  + \frac{k}{\kappa^2 g^2}(  \frac{f''}{f}   -\frac{f'g'}{fg}  + \frac{2f'}{rf}     )  &=&  YE^2   + \rho   \;, \label{gd1}\\
\frac{1}{\kappa^2 g^2}(
- \frac{2f'}{rf}   + \frac{g^2-1}{r^2}  ) +\frac{k}{\kappa^2 g^2}(\frac{f''}{f}   -\frac{f'g'}{fg}   - \frac{2g'}{rg}    ) &=& YE^2   - p  \;, \label{gd2}\\
\frac{1}{ \kappa^2 g^2}(   \frac{f''}{f} - \frac{f'g'}{fg}   + \frac{f'}{rf} -\frac{g'}{rg} )  + \frac{k}{ \kappa^2 g^2}(   \frac{g'}{rg} - \frac{f'}{rf} + \frac{g^2 -1 }{r^2}  )  &=& Y E^2   + p  \;, \label{gd3}
\end{eqnarray} 
under the condition (\ref{YRFk})
\begin{eqnarray}\label{cond2}
\frac{dY}{dR}  =  \frac{k}{ 2\kappa^2 E^2}  \;.
\end{eqnarray}
The conservation of the total energy-momentum tensor  for the gravitational field equation (\ref{gfe})  requires that 
\begin{eqnarray}\label{gd4}
p' +  (p+\rho) \frac{f'}{f}   - 2(YE)'E - \frac{4YE^2}{r} &=&0\; \;.
\end{eqnarray}

Assuming  the metric (\ref{metric}) has the conformal symmetry
$
L_\xi g_{ab} = \phi(r) g_{ab}
$  which describe the interior gravitational field  of  stars \cite{mak-harko-2004},\cite{herrera1,herrera2,herrera3},
the metric functions $f^2(r) $ and $g^2(r)$   were
found    in \cite{herrera1}  as
\begin{eqnarray}\label{s1}
f^2(r) &=& a^2 r^2, \hskip 3 cm 
g^2(r) = \frac{\phi^2_0}{\phi^2}= \frac{1}{X}.
\end{eqnarray}
Here $\phi(r)$ is an arbitrary  function of $r$, $ L_\xi $ is Lie derivative of the  metric tensor  along the vector field $\xi$, $X$ is a new function,  $  \phi_0 $ and $a$  are arbitrary  constants. Then     we obtain  the following  differential equation system  from (\ref{gd1})-(\ref{gd4}) under the symmetry,
\begin{eqnarray}
k(\frac{X'}{2\kappa^2 r} + \frac{2X }{\kappa^2 r^2}  )   -\frac{X'}{\kappa^2 r} +  \frac{ 1-X }{\kappa^2 r^2}    &=&   YE^2 + \rho\;, \label{d1}
\\
\frac{3k X'}{2 \kappa^2 r} +  \frac{1- 3 X}{\kappa^2 r^2}   &=&  Y E^2 -p   \;,\label{d2}\\
k(-\frac{X'}{2\kappa^2 r } + \frac{1-2X}{\kappa^2 r^2} ) + \frac{X'}{\kappa^2r } + \frac{X}{\kappa^2 r^2} &=& YE^2 +p \;,
\label{d3}\\
p' +  \frac{p}{r}  +\frac{\rho}{r} - 2(YE)'E - \frac{4YE^2}{r} &=&0\;. \label{d4}
\end{eqnarray}

with the condition (\ref{cond2}). Since we aim to extend the solution given in \cite{Sert2017}   to compact stars without introducing an equation of state,
   we  take  the following metric function with  the real numbers $\alpha> 2$ and   $b\neq 0$ inspired by \cite{mak-harko-2004},
\begin{eqnarray}\label{g1}
g^2(r) = \frac{1}{X} = \frac{3}{ 1 + br^\alpha} \;.
\end{eqnarray}
We see that the metric function is regular at center of coordinate $r=0$   and leads to  the following  regular  Ricci  scalar
\begin{eqnarray} \label{R}
R= - b(\alpha+2) r^{\alpha-2}.
\end{eqnarray}
Then we found  the following  class of   solutions   to  the  system of equation  (\ref{d1}-\ref{g1}) 
\begin{eqnarray}
E^2(r) &=& \frac{(1+k)
	(  br^\alpha(\alpha-2) +1)^{\frac{k(4\alpha+6) +\alpha}{(1+k) \alpha}}  }{6\kappa^2 r^2} \label{E1}
\;, \\
p(r) &=&  \frac{1+ k}{6\kappa^2 r^2} - \frac{br^\alpha ( 2\alpha k -\alpha +2k -4 )}{6\kappa^2 r^2} \;, \label{p1}\\
\rho(r) &=&  \frac{1+ k}{2\kappa^2 r^2} + \frac{br^\alpha ( 2k-\alpha )}{2\kappa^2 r^2} \;, \label{rho1} \\
Y(r) &=&  C\left[  1 +   b(\alpha-2)r^\alpha   \right]^{-\frac{3(\alpha+2)k}{(1+k)\alpha}} \;. \label{Y1}
\end{eqnarray}
where    $C $ is an integration constant. We see that the solutions are dependent on the parameter $k$ introduced by  (\ref{cond2}) together with the parameter  $\alpha$ in the metric function (\ref{g1}).  The charge of the star (\ref{q1}) inside  a spherical volume  with radius $r$ can be calculated by the charge-radius equality     
\begin{eqnarray}\label{qr1}
q^2(r) =   \frac{  (1+k)r^2 \left[1+b(\alpha-2)r^\alpha \right]^{-\frac{2\alpha k- \alpha + 6 k }{(1+k)\alpha} }}{6\kappa^2} \;.
\end{eqnarray}
It can be seen that the charge  is regular at   $r=0$    for $\alpha>2$.
By solving $r$
   in terms of $R$ as the inverse function $r(R)$ from (\ref{R}),  we re-express the  non-minimal coupling function   (\ref{Y1}) as
\begin{eqnarray}\label{Y2}
Y(R)= C \left[ 1 + b(\alpha-2)  (    \frac{-R}{\alpha b +2b}  )^{\frac{\alpha}{\alpha-2}}    \right]^{-\frac{3(\alpha+2)k}{(1+k)\alpha}}\;.
\end{eqnarray} 

 We can consider that the exterior vacuum region is described by Reissner-Nordsrom metric with $R=0$. 
 %But  when we consider the modified gravity models to explain dark matter and dark energy at large scale. However it is possible to ignore the high order corrections to the exterior Reissner-Nordstrom metric at the minimum level.
Then  the nonminimal function becomes  $Y(R) = C$ and our model involves   
   the    Einstein-Maxwell theory  as a minimal case at the exterior region with $C=1$.  
Thus we have  the more general Lagrangian of the  non-minimal   theory  which  describe   the compact stars for $\alpha>2\;$
\begin{eqnarray}\label{theory}
L =    
\frac{1}{2\kappa^2} R*1
-   \left[ 1+ b(\alpha-2)  (   \frac{-R}{\alpha b +2b}  )^{\frac{\alpha}{\alpha-2} }     \right]^{-\frac{3(\alpha+2)k}{(1+k)\alpha}} F\wedge *F  + 2A\wedge J + L_{mat}  + \lambda_a\wedge T^a \;. \ \ \ \ \ \ 
\end{eqnarray} 
The   field equations of the  Lagrangian (\ref{theory})   accept the solutions with the   electric field (\ref{E1}), pressure (\ref{p1}),
 energy density (\ref{rho1}), 
  electric charge (\ref{qr1})
and  the following metric tensor in the star
\begin{eqnarray}\label{metricin}
ds^2 _{in}= - a^2r^2dt^2 + \frac{3}{1+br^\alpha} dr^2 + r^2 d\Omega^2 \;. \end{eqnarray}

We will determine  the power $\alpha$ and $k$  in the model from   observational data and  the constant  $b$    from the matching and continuity conditions  (\ref{prs}).
In the absence of    the electromagnetic source and matter,  the Lagrangian of the non-minimally coupled theory (\ref{theory})  can reduce to the  Einstein-Maxwell Lagrangian  with $Y(R)=1$   
 at the exterior of the star
as a vacuum  case, and field equations of the minimal theory accept   
the well known  Reissner-Nordstrom metric 
\begin{eqnarray}\label{metricext}
ds^2_{out} = -(1-\frac{2M}{r} + \frac{\kappa^2Q^2}{r^2})dt^2 + (1-\frac{2M}{r} + \frac{\kappa^2Q^2}{r^2})^{-1}dr^2 + r^2d\Omega^2 \; 
\end{eqnarray}

with the exterior electric field 
$  E(r) = \frac{Q}{ r^2}$, where
$Q$ is     total electric charge of the star.

\section{Continuity Conditions}
The continuity of the interior  (\ref{metricin}) and exterior Reissner-Nordstrom   metric    
at the boundary  of the charged star   $r=r_b$    leads to
\begin{eqnarray}
a^2 = \frac{\kappa^2 Q^2   - 2Mr_b  + r_b^2  }{r_b^4}\;,\\
b =\frac{ 2r_b^2  - 6Mr_b  + 3\kappa^2Q^2  }{r_b^{2+\alpha}}\;. \label{b11}
\end{eqnarray}
At the boundary of the star the pressure (\ref{p1}) should be zero 
\begin{figure}[t]{}
	\centering
	\subfloat[ Pressure $  \kappa^2 p(r) = \frac{8\pi G}{c^4} p(r)  \ km^{-2}  $ ]{ \includegraphics[width=0.5\textwidth]{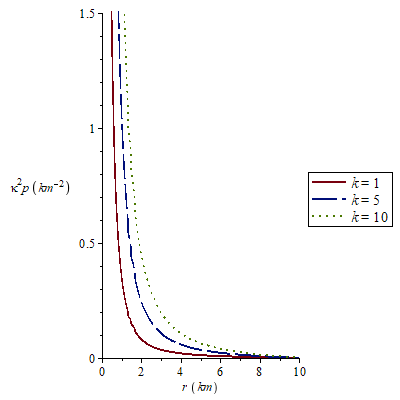} }
	\subfloat[ Energy density $\kappa^2c^2\rho(r)  = \frac{8\pi G }{c^2} \rho(r) \  km^{-2} $  ]{ \includegraphics[width=0.5\textwidth]{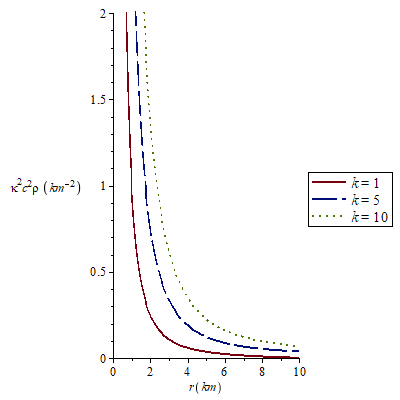}} 
	\\
	\parbox{6in}{\caption{{{\small{  Variation of  the pressure (a) and energy density (b) as a function of the radial distance $r$ inside the star with the  boundary radius $r_b=10 \; km$  and the parameter $\alpha=3$.
	}}}}}
\end{figure} 
\begin{eqnarray}
p(r_b) &=&  \frac{1+ k}{6\kappa^2 r_b^2} - \frac{br_b^\alpha ( 2\alpha k -\alpha +2k -4 )}{6\kappa^2 r_b^2}  
= 0   \;, \label{prs}
\end{eqnarray}
the condition determines the constant   $b$  in the non-minimal theory  (\ref{theory})   as
\begin{eqnarray}
b = \frac{1+k}{(2\alpha k- \alpha +2k -4 )r_b^\alpha}
\label{b1}\;.
\end{eqnarray}
Using this constant,  variation  of the pressure and energy density  as a function of the radial distance $r$ inside the star is given in Fig. 1.
The interior of  the star is considered  as the specific fluid   which has very high  gravitational fields,  electromagnetic  fields and matter. Then  the electromagnetic fields  obey  the modified Maxwell field equation $d*YF =J $ in matter. The integral of the modified Maxwell equation (\ref{maxwell1}) gives  the  charge   in   volume  with radius $r$  in the star,
$q(r) = YEr^2 $ (\ref{qr1}). On the other hand,  the Ricci scalar is zero at the exterior of the star  and then,  using  (\ref{Y2}), we can take  the  non-minimal function as   $Y=1$   and  obtain the Maxwell field equation $d*F=0$ which leads to the 
   the total charge  $Q= E r^2$ at the exterior  region.
Here    the displacement field $YE$  inside of the star turns out to be  the electric field $E$  outside. Setting by $r=r_b$   in eq. (\ref{qr1})
the total charge of the star $Q=q(r_b)$ is found as 
\begin{eqnarray}\label{Q1}
Q^2  =  \frac{  (1+k)r_b^2 \left[1+    \frac{(1+k)(\alpha-2)}{(2\alpha k- \alpha +2k -4 )}  \right]^{-\frac{2\alpha k- \alpha + 6 k }{(1+k)\alpha} }}{6\kappa^2}  \;
.
\end{eqnarray} 
Then the outside electric field  is given by
$E=Q/r^2$.
The total charge-boundary radius ratio obtained from  (\ref{Q1}) is plotted by Figure 2a dependent on the parameter $\alpha$  for some different $k$ values.

\begin{figure}[t]{}
	\centering
	\subfloat[ $ \frac{8\pi G}{c^4} \frac{ Q^2}{r_b^2} $ versus $\alpha$ ]{ \includegraphics[width=0.5\textwidth]{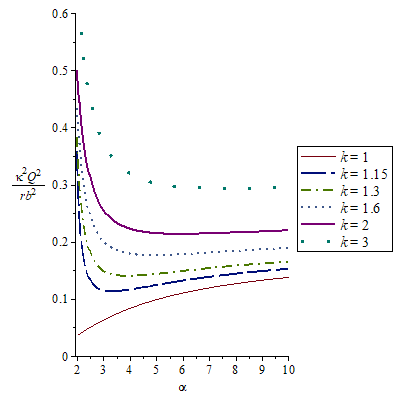} }
	\subfloat[ $\frac{GM_m}{c^2 r_b} $ versus $\alpha$ ]{ \includegraphics[width=0.5\textwidth]{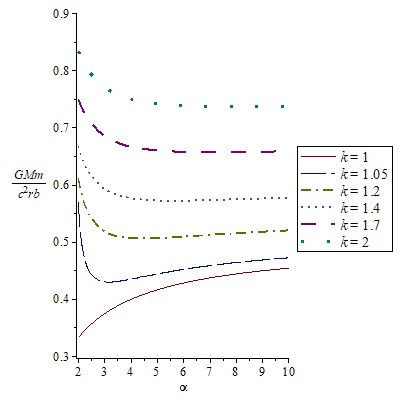}} 
	\\
	\parbox{6in}{\caption{{{\small{ Variation of the dimensionless  quantities which are related with  the total  electric  charge  $Q$  and  the matter mass  $M_m$  as a function of     the dimensionless parameter    $\alpha$ for some different   $k$ values.
	}}}}}
\end{figure} 
The matter mass of the  star   $M_m$ can be found from the integral of the energy density  $\rho$ 
\begin{eqnarray}\label{Mm1}
M_m = \frac{\kappa^2}{2} \int_0^{r_b}  \rho r^2 dr =   \frac{r_b (1+k) ((k-\frac{1}{2})\alpha^2 + (2k-3)\alpha + 2k -2 )}{2(2\alpha k -\alpha +2k -4)(\alpha+1)}\;.
\end{eqnarray}
The figure of the matter mass $M_m$    is shown in  Figure 2b  for increasing  $\alpha $ and some different $k$ values.  We see that   the total charge  and matter  mass of the star  increase with the increasing $k$ values.

When we compare the constant  $b$ given by the equations  (\ref{b1}) and (\ref{b11}), we find 
the following mass-charge relation for the model
\begin{eqnarray}
\frac{M}{r_b}= \frac{4\alpha k -2\alpha +3k-9}{6(2\alpha k -\alpha + 2k -4 )} + \frac{\kappa^2 Q^2}{2r_b^2} \; \label{M2} \;.
\end{eqnarray} 
By substituting  the total charge   (\ref{Q1}) in 
(\ref{M2}),   the total mass-radius ratio $M/r_b$ can be found  in terms of the parameters $\alpha$ and $k$ of the model 
\begin{eqnarray}
\frac{M}{r_b}= \frac{4\alpha k -2\alpha +3k-9}{6(2\alpha k -\alpha + 2k -4 )} + \frac{1+k}{12} [1+ \frac{(1+k)(\alpha-2)}{2\alpha k -\alpha +2k -4 }]^{-\frac{\alpha (2k-1) +6k}{(1+k)\alpha}}  \; \label{M1} 
\;. \label{M3}
\end{eqnarray}
The   gravitational  redshift $z$  at the boundary  is obtained  from
\begin{eqnarray}\label{z}
z= (1-\frac{2M}{r_b} + \frac{\kappa^2 Q^2}{r_b^2})^{-
	\frac{1}{2}} -1  = \sqrt{\frac{3(2\alpha k -\alpha +2 k -4)}{2\alpha k -\alpha +3k -3} } -1 \;.
\end{eqnarray}

By  taking the limit $\alpha ,k \rightarrow \infty$,
the maximum  redshift is found as
$z = \sqrt{3} -1 \approx 0.732$ same  with $k=1$ in \cite{Sert2017}. The upper redshift bound    is smaller than    the Buchdahl bound $z = 2$  and the bound   given in \cite{Mak1}.

The mass-radius relation and the gravitational redshift-radius relation are shown in Figure 3a and 3b, respectively,   depending on the    parameter $\alpha$ for some $k$ values. 	We see that as the k value increases, the mass and redshift  increases.

\begin{figure}[h]{}
	\centering
	\subfloat[ 
	$\frac{GM}{c^2 r_b} $   ]{ \includegraphics[width=0.5\textwidth]{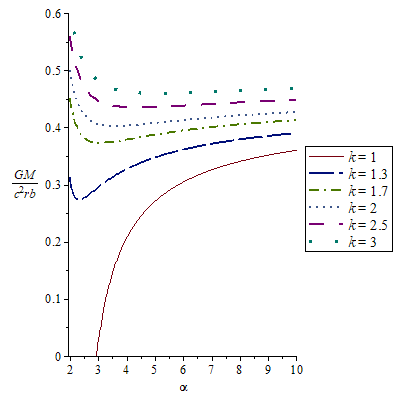} }
	\subfloat[ Redshift $z$ ]{ \includegraphics[width=0.5\textwidth]{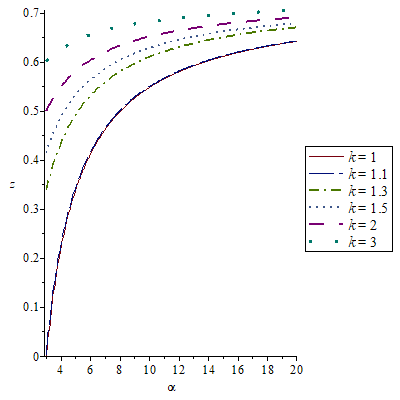}} 
	\\
	\parbox{6in}{\caption{{{\small{ 
	Variation of the dimensionless  quantity which is related with  the gravitational mass $M$ and     the gravitational surface redshift  $z$  as a function of      the dimensionless parameter $\alpha$ for some different   $k$ values. 
	}}}}}
\end{figure}

To obtain an interval for the parameter $k$ 
we consider  the energy density  condition inside the  star
\begin{eqnarray}
\rho(r)  = \frac{1+ k}{2 \kappa^2 r^2} -\frac{b(\alpha-2k)r^{(\alpha-2)}   }{2 \kappa^2 } \geq 0  \label{rho2} \;.
\end{eqnarray}
At the center of the star  the condition 
\begin{eqnarray}
\lim\limits_{r\rightarrow 0} \rho(r) =  \frac{1+k}{2\kappa^2 r^2} \geq 0
\end{eqnarray}
gives  $k\geq -1 $. On the other hand, at the boundary $r=r_b$
the condition (\ref{rho2}) turns out  to be
\begin{eqnarray}\label{ineq3}
1 \geq \frac{(\alpha-2k)}{2\alpha k -\alpha +2 k -4}\;.
\end{eqnarray}
The solution of the inequality (\ref{ineq3}) is
\begin{eqnarray}\label{ineq1}
k \geq 1 \hskip 1 cm or \hskip 1 cm -1<  k   \leq \frac{\alpha+4}{2\alpha+2}\;.
\end{eqnarray}

%On the other hand, if we consider  the positive gravitational surface redshift $z>0$  in  equation (\ref{z}), we obtain
%\begin{eqnarray}\label{ineq2} k> \frac{2\alpha +9}{4\alpha +3} \hskip 1 cm or \hskip 1 cm k < \frac{\alpha+3}{2 \alpha +3}\;. \end{eqnarray}
Then we can choose    $k\geq 1$ without loss of generality.
The derivative of the pressure $p(r)$ according to the energy density $\rho(r)$ is calculated as
\begin{eqnarray}
\frac{dp}{d\rho }  =   \frac{(2\alpha k - \alpha +2k -4 ) (\alpha r^\alpha +2 r_b^\alpha -2 r^\alpha)}{ 6(2\alpha k - \alpha +2k -4)r_b^\alpha  + 3(\alpha-2k)(\alpha-2)r^\alpha }\;.
\end{eqnarray} 
The  phase speed   of  the sound waves   in the star is defined by $  (\frac{dp}{d\rho } )^{1/2} $ and  the speed   satisfies  the causality  condition  $  (\frac{dp}{d\rho } )^{1/2} <1 $  for  $k\geq 1$ and $\alpha>2 $ values (see Fig. 6a for some values), where
the speed of light $ \mathtt{c} =1$. 
Thus each possible values of the parameters $k>1$ and $\alpha>2$ in the modified  model give a mass, charge-radius ratio   and  redshift $z$.

\subsection{The simple model with $\alpha=3 $}

We consider the simple case setting  by $\alpha= 3$  in the non-minimally coupled model  (\ref{theory})
\begin{eqnarray}\label{theory3}
L =    
\frac{1}{2\kappa^2} R*1
-    \left[ 1 -     \frac{R^3}{(\alpha +2)^3 b^2}      \right]^{-\frac{ 5 k}{(1+k) }} F\w *F  + 2A\wedge J + L_{mat}  + \lambda_a\wedge T^a\;. \ \ \ \ \ \ 
\end{eqnarray} 
Here we note that the non-minimal coupling  function $Y(R)$ can be expanded as the binomial series in power of $R^3$ for $\abs{   \frac{R^3}{(\alpha +2)^3 b^2}  } < 1$.
\begin{eqnarray}
Y(R) = \left[ 1 -     \frac{R^3}{(\alpha +2)^3 b^2}      \right]^{-\frac{ 5 k}{1+k }} = 1 + \frac{5k}{1+k}  \frac{R^3}{(\alpha +2)^3 b^2}   + \mathcal{O}(R^6)
\end{eqnarray}
  The non-minimal model (\ref{theory3})
accept the interior metric as a solution
\begin{eqnarray}\label{metricin3}
ds^2 = - a^2r^2dt^2 + \frac{3}{1+br^3} dr^2 + r^2(d\theta^2 +sin^2\theta d\phi^2 )
\end{eqnarray}
together with the following electric field, pressure, energy density and electric charge  from (\ref{E1})-(\ref{qr1}) inside of the star
\begin{eqnarray}
E^2(r) &=& \frac{(1+k)
	(  br^3 +1)^{\frac{1+6k}{1+k}}  }{6\kappa^2 r^2} \label{rho3}
\;, \\
p(r) &=&  \frac{1+ k}{6\kappa^2 r^2} + \frac{br ( 8 k  -7 )}{6\kappa^2}\;, \label{p4}\\
\rho(r) &=&  \frac{1+ k}{2\kappa^2 r^2} + \frac{br ( 2k-3 )}{2\kappa^2 }\;,  \label{rho4} \\
q^2(r) &=&   \frac{  (1+k)r^2 \left[1+br^3 \right]^{-\frac{4 k-1  }{(k+1)} }}{6\kappa^2} \;.
\end{eqnarray}

 Then under  the boundary and matching conditions    the  total  charge, mass and gravitational redshift   can be expressed   by the followings

\begin{eqnarray}\label{Q3}
Q^2  =  \frac{  (1+k)r_b^2 \left[1+    \frac{(1+k) }{(8 k- 7 )}  \right]^{-\frac{4 k-1 }{(1+k)} }}{6\kappa^2}  \;,
\end{eqnarray} 
\begin{eqnarray}
\frac{M}{r_b}= \frac{15k-15}{6(8 k -7)} + \frac{1+k}{12} [1+ \frac{(1+k)}{8 k -7}]^{-\frac{4 k-1 }{(1+k)} }  \;, \label{M4} 
\; 
\end{eqnarray}
\begin{eqnarray}\label{z3}
z=   \sqrt{\frac{3(8k-7)}{9k -6 } } - 1 \;.
\end{eqnarray}
from (\ref{Q1})-(\ref{z}).
We note that the redshift has the upper bound $z\approx 0.633 $  for $\alpha=3$, which    is smaller than the  general  redshift  bound of the model which is $\sqrt{3} -1 \approx 0.732$. As the simple model  with  $\alpha =3 $, we depict the related physical  quantities in Fig 4-6. We  give the corresponding  $k$ values  
by taking   the   observed mass-radius ratios of  some known stars  and calculate the other quantities in Table 1.
Here we note that each star  have its own $k$ value and the each $k$ value can be determined by the  observational mass and radius of the star.

\begin{table}[t]
	{\small 
		\centering
		\begin{tabular}{|c|c|c|c|c|}
	\hline  \ \
		 Star   & $\frac{M}{r_b} \ (\frac{M_{\odot}}{km})$  &$k $ &  $\frac{\kappa^2Q^2}{r_b^2}$ &
		 $z$ (redshift) \\
	\hline	EXO 1745-248  & $\frac{1.4}{11} $ \cite{Ozel2009}
	&  1.047  & 0.082 & 0.098 \\
			\hline
	4U 1820-30  &    $\frac{1.58}{9.11}$ \cite{Guver2010} & 1.084 & 0.095 & 0.156 \\

	\hline			4U1608-52
	 &
	$\frac{1.74}{9.3}$ 
	\cite{Guver20102}  
		  & 1.098 &  0.099 & 0.174\\
			\hline			
SAX J1748.9-2021 &   
$\frac{1.78}{8.18}$ \cite{Guver2013} & 1.135 & 0.110 & 0.217\\
\hline
\end{tabular}
		\caption{
			The dimensionless  parameter $k$, the charge-radius ratio $\frac{\kappa^2Q^2}{r_b^2} $ and the surface redshift  $z$ obtained by using the observational mass $M$ and the radius $r_b$ for some neutron stars.	}
		\label{tab:template}
	}
\end{table}
In order to obtain an approximate equation of state  of the matter inside the star from equation (\ref{p4}) and (\ref{rho4}), we fit the parametric $p-\rho$ curve with the equation

\begin{eqnarray}\label{eos}
 \kappa^2 p=\frac{\kappa^2c^2\rho}{3} - \frac{k-1}{6}B_0 \  
\end{eqnarray}
where we have used  $B_0= 1 km^{-2}$ as a dimension-full constant.   We note that if $k=1$
we obtain the radiation fluid case $p=c^2\rho/3$ \cite{Sert2017}.
 The curve fitting is shown in Fig. 7.
 It is interesting to see that the fitting equation of state corresponds to the MIT bag model \cite{Witten,Cheng} which is given by  $p=\frac{ c^2 \rho}{3} - \frac{4 c^2B }{3} $.  In this model the bag constant $B$ is related to the parameter $k$ as $B= \frac{k-1}{8} \frac{km^{-2}}{c^2\kappa^2}   $.
 If we set  $k=1.015 $, 
we find $ B = 1.88 \times10^{-13}  \frac{cm^{-2}}{c^2\kappa^2}  = 10^{14}gr/cm^3   $,  
where we have used $\kappa^2 = \frac{8 \pi G}{c^4} =  2.1 \times 10^{-48} \frac{s^2}{gr \; cm}$ and $ c= 2.99 \times 10^{10} \frac{cm}{s}$.
Then $B < 10^{14}gr/cm^3$  for  $1< k< 1.015 $   and  $B > 10^{14}gr/cm^3$ for  $k> 1.015$. Additionally, we see that    the parameter $k$ takes  values  approximately in the interval $1<k<1.2$ from the Table 1 for realistic compact stars. In the case $\alpha=3$, the non-minimal Einstein-Maxwell model with $k=1.015$ gives the gravitational  mass $M=0.65 M_{\odot} $ using the radius $r_b =9.46 km $.  We see that  this mass  is  less   than   the mass obtained for conformal symmetric charged stellar models in the minimal Einstein-Maxwell theory \cite{mak-harko-2004}. We note that  the both models  have the conformal symmetry and  very similar interior metric solutions.
Although this modified Einstein-Maxwell model   has  free parameters  which  can be set to be consistent with various compact star observations,
the model in \cite{mak-harko-2004}
determines a unique charged configuration of quark matter  in terms of the Bag constant. 
Alternatively, the mass in our model increases to  $2.09M_\odot$ for $\alpha=4$, and  $2.64M_\odot$ for $\alpha=5$. That is,  we can  confine $k$ as $k=1.015$ which gives the Bag constant and consider  $\alpha $  is a  free parameter, in order to describe compact stars.

\section{Conclusion}
We have extended the solutions of  the previous  the non-minimally coupled $Y(R)F^2$ theory \cite{Sert2017} to the case with $\rho\neq 3p$ 
under the symmetry of conformal motions.
In the case without  any assumption of  the equation of state,  we have acquired one more parameter $k$  in the solutions and the corresponding model. The  pressure and energy density in the solutions      are decreasing function with $r$    in the interior of the star.
%We have determined  the interval of the parameter $k$  as $k\geq 1$ using  the   energy  density  and  the surface gravitational redshift to be positive, $\rho\geq 0$, $ z>0 $.

By matching the   interior  solution with the exterior Reissner-Nordstrom  solution
and 
 applying   the zero pressure condition at the boundary radius of star $r=r_b$,     we determine some physical properties of the star such as the ratio of the total mass and charge to boundary radius $r_b$ and gravitational redshift  depending on the parameters $k$ and $\alpha$. 

We note that the total mass and charge increase with increasing $ k $ values and we have not reach an upper bound for $k$ in the extended non-minimal model.
 But the increasing $k$ values give an upper bound for  the gravitational redshift,
$z \approx 0.732$, which is   smaller than  the  more general restriction  found in  \cite{Mak1} for compact  charged objects.
It is interesting to note that  each     $\alpha$ and $k$  value in this model    (\ref{theory}) determines a different  non-minimally coupled theory and each theory with the different parameters gives   different     mass-radius, charge-radius ratios  and gravitational redshift configuration.
    We calculated $k$ values and the corresponding other quantities of the compact stars with  the simple case $\alpha=3$   via    the some observed mass-radius values in Table 1. In this case, we also obtained the  approximate equation of state  (\ref{eos})  by fitting the $p-\rho$ curve of the model. By comparing the fitting equation of state with the MIT bag model for $\alpha=3$, we  found  the gravitational mass $M=0.65 M_\odot$ which is smaller than the mass obtained for conformal symmetric quark stars \cite{mak-harko-2004} in the Einstein-Maxwell theory. However  the mass in our model  increases  as $\alpha$ increases.
    The non-minimally coupled model has  the arbitrary parameters $\alpha$ and $k$ which can be set in order to be consistent with compact star observations. Even for $\alpha=3$, each $k$
    value  can describe  a charged  compact star.

%\section*{Acknowledgement}
%This study is supported by the Scientific Research Project (BAP) 2017HZDP009, Pamukkale  University, Denizli, Turkey.

%\newpage

\begin{figure}[h]{}
		\centering
		\subfloat[  ]{ \includegraphics[width=0.38\textwidth]{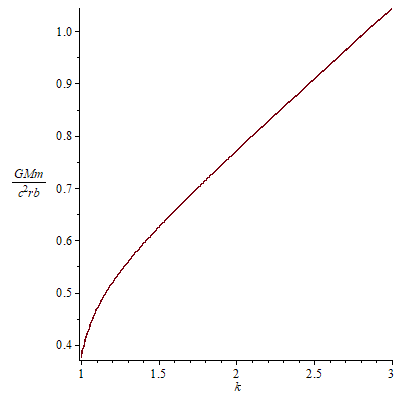} }
		\subfloat[    ]{ \includegraphics[width=0.38\textwidth]{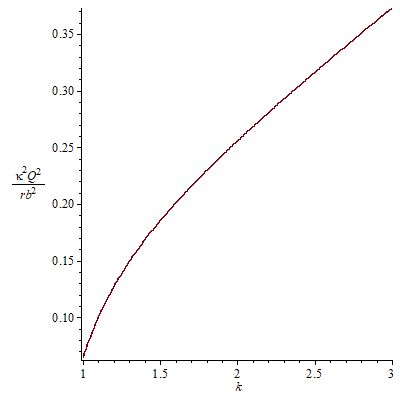}} 
		\\
		\parbox{6in}{\caption{{{\small{  The dimensionless  quantities which are related with   the matter mass  (a)     and    total  electric  charge  (b)  versus    the dimensionless parameter    $k$  for $\alpha = 3$. 
		}}}}}
	\end{figure}

	\begin{figure}[t]{}
		\centering
		\subfloat[   ]{ \includegraphics[width=0.38\textwidth]{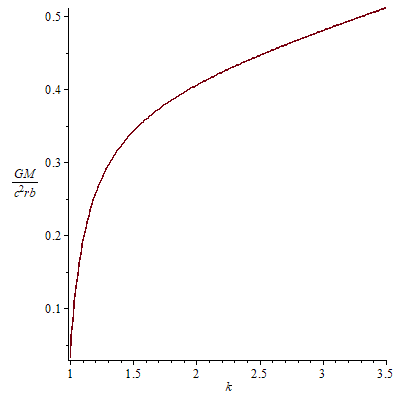} }
		\subfloat[   ]{ \includegraphics[width=0.38\textwidth]{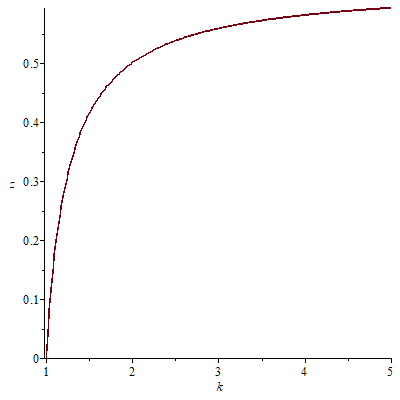}} 
		\\
		\parbox{6in}{\caption{{{\small{ 
							The dimensionless  quantity which is related with    the gravitational mass   (a)  and   the gravitational surface redshift (b)    versus    the dimensionless parameter    $k$  for $\alpha = 3$.   
		}}}}}
	\end{figure}
	
	\begin{figure}[t]{}
		\centering
		\subfloat[  $\alpha=3 $  ]{ \includegraphics[width=0.38\textwidth]{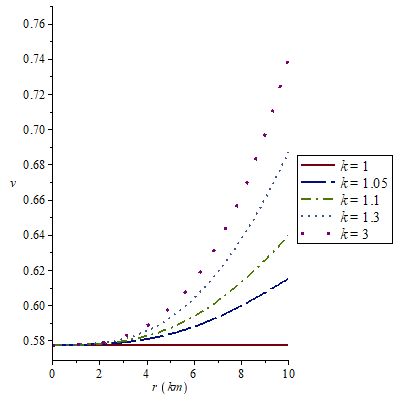} }
		\subfloat[ $\alpha=5 $  ]{ \includegraphics[width=0.38\textwidth]{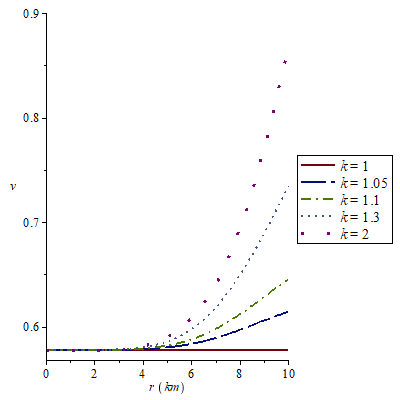}} 
		\\
		\parbox{6in}{\caption{{{\small{ 
	Variation of the phase speed $v=\frac{dp}{d\rho} $  as a function of radial distance  $r$  
using  the boundary radius $r_b=10 km$  and some $k$ values 	for $\alpha=3 $ (a) and   $\alpha=5 $ (b).
		}}}}}
	\end{figure}

	\begin{figure}[t]{}
		\centering
		\subfloat[    $k=1.015$] { \includegraphics[width=0.5\textwidth]{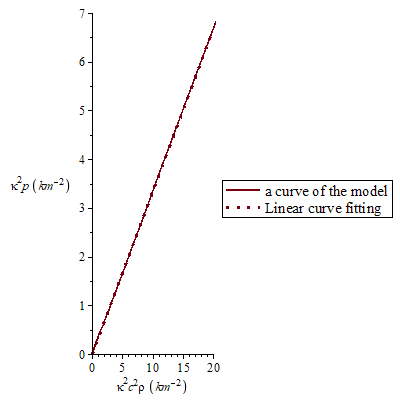} }
		\subfloat[  $ k=1.1$  ]{ \includegraphics[width=0.5\textwidth]{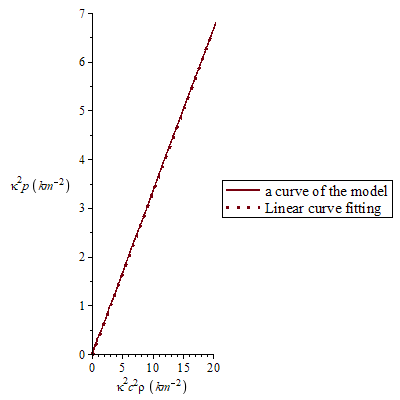}} 
		\\
		\parbox{6in}{\caption{{{\small{ 
							The pressure versus energy density inside the star for   $r_b=10\ km, $   and  $\alpha=3 $.
							The solid curves in (a) and (b) represent 	the relation between   $p$ and  $\rho$  from (\ref{p4}) and (\ref{rho4})  and the dotted curves represent  the equation of state $p=\frac{\rho}{3} - \frac{k-1}{6} $  by obtained from curve fitting. 
		}}}}}
	\end{figure}

\end{document}